# Structural and transport properties of highly Ru-deficient SrRu$_{0.7}$O$_3$ thin films prepared by molecular beam epitaxy: comparison with stoichiometric SrRuO$_3$

Yuki K. Wakabayashi,[1,*,†] Shingo Kaneta-Takada,[1,2,*] Yoshiharu Krockenberger,[1] Kosuke Takiguchi,[1,2] Shinobu Ohya,[2,3] Masaaki Tanaka,[2,4] Yoshitaka Taniyasu,[1] and Hideki Yamamoto[1]

[1]NTT Basic Research Laboratories, NTT Corporation, Atsugi, Kanagawa 243-0198, Japan
[2]Department of Electrical Engineering and Information Systems, The University of Tokyo, Bunkyo, Tokyo 113-8656, Japan
[3]Institute of Engineering Innovation, The University of Tokyo, Bunkyo, Tokyo 113-8656, Japan
[4]Center for Spintronics Research Network (CSRN), The University of Tokyo, Bunkyo, Tokyo 113-8656, Japan

[*]These authors contributed equally to this work.
[†]Author to whom correspondence should be addressed: yuuki.wakabayashi.we@hco.ntt.co.jp

Abstract

We investigate structural and transport properties of highly Ru-deficient SrRu$_{0.7}$O$_3$ thin films prepared by molecular beam epitaxy on (001) SrTiO$_3$ substrates. To distinguish the influence of the two types of disorders in the films—Ru vacancies within lattices and disorders near the interface—SrRu$_{0.7}$O$_3$ thin films with various thicknesses ($t$ = 1–60 nm) were prepared. It was found that the influence of the former dominates the electrical and magnetic properties when $t \geq 5-10$ nm, while that of the latter does when $t \leq 5-10$ nm. Structural characterizations revealed that the crystallinity, in terms of the Sr and O sublattices, of SrRu$_{0.7}$O$_3$ thin films, is as high as that of the ultrahigh-quality SrRuO$_3$ ones. The Curie temperature ($T_C$) analysis elucidated that SrRu$_{0.7}$O$_3$ ($T_C \approx 140$ K) is a material distinct from SrRuO$_3$ ($T_C \approx 150$ K). Despite the large Ru deficiency (~30%), the SrRu$_{0.7}$O$_3$ films showed metallic conduction when $t \geq 5$ nm. In high-field magnetoresistance measurements, the fascinating phenomenon of Weyl fermion transport was not observed for the SrRu$_{0.7}$O$_3$ thin films irrespective of thickness, which is in contrast to the stoichiometric SrRuO$_3$ films. The (magneto)transport properties suggest that a picture of carrier scattering due to the Ru vacancies is appropriate for SrRu$_{0.7}$O$_3$, and also that proper stoichiometry control is a prerequisite to utilizing the full potential of SrRuO$_3$ as a magnetic Weyl semimetal and two-dimensional spin-polarized system. Nevertheless, the large tolerance in Ru composition (~30 %) to metallic conduction is advantageous for some practical applications where SrRu$_{1-x}$O$_3$ is exploited as an epitaxial conducting layer.

The itinerant 4$d$ ferromagnetic perovskite SrRuO$_3$ has attracted strong attention because of the unique nature of its ferromagnetism, metallicity, chemical stability, and compatibility with other perovskite-structured oxides.[1–17] It has been widely used in oxide electronics and spintronics as an epitaxial conducting layer.[8] Recently, interest in SrRuO$_3$ has been further boosted by the observation of Weyl fermions[18,19] and the realization of two-dimensional ferromagnetism in electrically conducting layers of one-unit-cell thickness embedded in [(SrRuO$_3$)$_1$/(SrTiO$_3$)$_n$] heterostructures.[20,21] These intriguing phenomena are observed only in samples of exceptionally high quality;[18,20] specifically, those with a high residual resistivity ratio (RRR), which is known to be a good indicator of the purity of metallic systems.[8,11,18]

To apply the magnetic Weyl semimetal state and two-dimensional ferromagnetism in SrRuO$_3$-based heterostructures to newly proposed spintronic devices[22] and topoelectrical circuits,[23,24] we must be able to control the interfaces between the SrRuO$_3$ and other layers. A promising approach to gain insight into the interfaces is to investigate thickness-dependent (magneto)transport properties of SrRuO$_3$ films, especially down to the nanometer scale. This is because the contribution of interface-driven disorders on the transport properties varies with thickness.[19] Several studies on such ultra-thin SrRuO$_3$ films, including our previous ones, have demonstrated higher resistivity and a lower Curie temperature ($T_C$) with decreasing film thickness.[8,25–29] When SrRuO$_3$ film thickness is below 2 nm, neither residual resistivity (thus, the RRR) nor $T_C$ could be well-defined because an insulating and nonferromagnetic state emerges.[25–29] The only exception is atomically thin SrRuO$_3$ (0.4 nm) layers embedded in [(SrRuO$_3$)$_1$/(SrTiO$_3$)$_n$] heterostructures that show ferromagnetic and conducting behavior.[20,21] On the other hand, our recent study showed that the threshold RRR to observe transport phenomena stemming from the Weyl fermions is ~20, which corresponds to thickness $t \geq 10$ nm for stoichiometric films reasonably free from Ru vacancies.[19]

Here, there are two possible origins of low RRR values in SrRuO$_3$. One is poor crystallinity due to high impurity density and/or structural disorders, which is generally seen in low-quality samples.[8,11,12,30,31] The other one is Ru vacancies, which is rather specific to SrRuO$_3$.[8,11] It is therefore required to distinguish between these two possibilities in order to improve the quality of SrRuO$_3$ samples. Stringent stoichiometry control during the film growth is required if the the Ru vacancies dominate disorder. However, those two sources have sometimes been mixed up, and there is a dearth of knowledge on the influence of off-stoichiometry on the transport, especially quantum transport properties.

In this study, we investigated the (magneto)transport properties of epitaxial SrRuO$_3$ films with various thicknesses grown under highly Ru-deficient conditions, and compared the results with those for stoichiometric films, which we have already reported.[18,19] Using molecular beam epitaxy (MBE), we can vary the ratio of the Sr and Ru beam fluxes supplied *ad arbitrium* during the growth and thus design and prepare such highly Ru-deficient SrRu$_{1-x}$O$_3$ thin films whose crystallinity, in terms of the Sr and O sub-lattices, is as high as that of stoichiometric (SrRuO$_3$) thin films. Similar composition controls are almost impossible with other thin-film growth techniques such as pulsed-laser deposition and sputtering, where the preparation of off-stoichiometric targets is a prerequisite. The chemical composition of a Ru-deficient film with thickness $t = 60$ nm was determined using energy dispersive X-ray spectroscopy (EDS) measurements. Complementary information was obtained by X-ray photoelectron spectroscopy (XPS). For comparison, the EDS and XPS measurements were also carried out for a stoichiometric film with identical thickness.

Ru-deficient SrRu$_{1-x}$O$_3$ films with thickness $t$ ranging from 1 to 60 nm were grown on (001) SrTiO$_3$ substrates in a custom-designed MBE setup equipped with multiple e-beam evaporators for Sr and Ru.[18, 33] For comparison, a 60-nm-thick stoichiometric SrRuO$_3$ film was also prepared. Detailed information about our MBE setup and preparation of the substrates are described elsewhere.[34–36] Oxidation during the growth was carried out with a mixture of ozone (O$_3$) and O$_2$ gas (~15% O$_3$ + 85% O$_2$), which was introduced through an alumina nozzle pointed at the substrate at a flow rate of ~2 sccm. The growth temperature was fixed at 772°C for all the films.

We precisely controlled the elemental fluxes, even for elements with high melting points, e.g., Ru (2250°C), by monitoring the flux rates with an electron-impact-emission-spectroscopy sensor, which were fed back to the power supplies for the e-beam evaporators. The Ru flux rates were set at 0.190 and 0.365 Å/s for the growth of the Ru-deficient and stoichiometric films, respectively, while the Sr flux was kept at 0.98 Å/s. The growth rate of 1.05 Å/s was deduced from the thickness calibration of a thick (60 nm) SrRuO$_3$ film using cross-sectional scanning transmission electron microscopy (STEM). This growth rate agrees very well with that (1.08 Å/s) estimated from the flux rate of Sr, confirming the accuracy of the film thickness. This also indicates that the sticking coefficient of Sr can be deemed to be unity. In contrast, it is known that supplied Ru partially re-evaporates from the growth surface through the formation of volatile species such as RuO$_4$ and RuO$_3$ under oxidizing atmosphere,[10,11] especially when excessive Ru is supplied; hence, one needs to grow films under a Ru-rich condition to form stoichiometric ones. The supplied Ru rates of 0.190 Å/s for Ru-deficient and 0.365 Å/s for stoichiometric films correspond to the Ru/Sr flux ratios of 0.80 (Ru-poor) and 1.54 (Ru-rich), respectively.

To determine the Ru/Sr composition ratios in the grown films, the 60-nm-thick films prepared under the Ru-poor and Ru-rich conditions were measured by EDS using a Bruker Quantax-400 EDS spectrometer [Fig. 1(a)]. The energy of the incident electron beam was 5 keV, and the probing area was 85 × 115 $\mu$m$^2$. Only the Sr, Ru, and O peaks are observed, indicating that the signals are exclusively from the films and the contribution from the SrTiO$_3$ substrates is negligible. We estimated the chemical compositions of the films using the conventional ZAF matrix correction routine built into the Bruker Esprit software,[37,38] where Z, A, and F represent atomic number, absorption, and fluorescence corrections, respectively. The results indicate that the Ru/Sr ratios in the films grown under the Ru-poor and Ru-rich conditions are 0.71 and 1.03, respectively. The Ru/Sr ratio of 0.71 in the Ru-deficient film seems to be reasonable considering the upper limit set by the supplied Ru/Sr flux ratio is 0.80. On the other hand, the film grown under the Ru-rich condition turned out to be stoichiometric within the accuracy of EDS. It is possible that O sites are also vacant to some extent in the Ru-deficient films. Although quantitative estimation of oxygen content by EDS is elusive, a less intense O peak for the Ru-deficient film than the stoichiometric one [Fig. 1(a)] supports this possibility. Under the naive assumption that the O content is proportional to the peak intensity and that the stoichiometric film has a composition formula of SrRuO$_3$, the Ru-deficient film is expressed as SrRu$_{0.71}$O$_{2.87}$. When the chemical formula of the Ru-deficient film is SrRu$_{0.7}$O$_{2.9}$ or SrRu$_{0.7}$O$_3$, the average valence of Ru is evaluated to be

+5.4 ($Ru^{5.4+}$) or +5.7 ($Ru^{5.7+}$), respectively, which is much higher than that (+4) in stoichiometric $SrRuO_3$ ($Ru^{4+}$). The existence of such a high valence state is suggested by the chemical shift in the Ru $3p_{3/2}$ XPS spectra [Fig. 1(b)]:[39-42] the Ru $3p_{3/2}$ level has slightly higher binding energy in $SrRu_{0.7}O_3$ (463.7 eV) than in $SrRuO_3$ (463.4 eV). The lower peak intensity in the XPS spectra for the $SrRu_{0.7}O_3$ film is also, at least qualitatively, consistent with the Ru/Sr ratio analysis by EDX. Those XPS measurements were performed with an ULVAC-PHI Model XPS5700 with a monochromatized Al Kα (1486.6 eV) source operated at 200 W. Hereafter, the Ru-deficient films are denoted by $SrRu_{0.7}O_3$ because our main interest is the Ru-deficiencies and because reliable estimation of the O content by EDX or XPS is impossible.

Figure 2(a) shows the RHEED patterns of the Ru-deficient $SrRu_{0.7}O_3$ surfaces for various thicknesses $t$. Every film shows sharp streaky patterns, indicating that the growth of the $SrRu_{0.7}O_3$ film proceeded in a two-dimensional layer-by-layer manner. Figure 2(b) shows cross-sectional high-angle annular dark-field scanning transmission electron microscopy (HAADF-STEM) images of a Ru-deficient film with $t = 60$ nm. The $SrRu_{0.7}O_3$ film was grown epitaxially on a (001) $SrTiO_3$ substrate with an abrupt substrate/film interface, as expected from the RHEED patterns. Unexpectedly, these RHEED patterns and STEM images for the Ru-deficient $SrRu_{0.7}O_3$ films are almost identical to those for the stoichiometric $SrRuO_3$ films reported previously,[19] even though around 30% of Ru sites are missing. This indicates that Ru vacancies are not line or plane defects but point defects, which are insensitive to RHEED and cross-sectional STEM measurements when they are randomly distributed. The caveat here is that $SrRu_{0.7}O_3$ is indistinguishable from $SrRuO_3$ in HAADF-STEM, which has atomic resolution as well as elemental discrimination capability. In Fig. 2(c), $\theta$-$2\theta$ X-ray diffraction (XRD) patterns are compared between the $SrRu_{0.7}O_3$ and $SrRuO_3$ films with a common thickness of 60 nm. The sharpness, intensity, and relative intensity of the peaks [not shown in Fig. 2(c)] are essentially identical between the two XRD patterns. In addition, Laue fringes around the (004) peaks (peaks are indexed for pseudocubic lattices) are clearly observed in both patterns. These results indicate that the overall crystallinity of the $SrRu_{0.7}O_3$ film is as high as that of the $SrRuO_3$ film. The out-of-plane lattice constant estimated by the Nelson-Riley extrapolation method is 3.951 Å for the Ru-deficient $SrRu_{0.7}O_3$ film, which is slightly larger than the value (3.949 Å) for the stoichiometric $SrRuO_3$ film.[19] This is consistent with previous reports claiming that the existence of Ru vacancies increases the out-of-plane lattice constant of $SrRuO_3$ films on $SrTiO_3$ substrates.[11,43] While preparation of bulk single crystals with highly Ru-deficient compositions (~30%) remains to be done, we speculate that epitaxial stabilization may have fostered the formation of the perovskite-structured $SrRu_{0.7}O_3$ films whose crystallinity is comparably high to that of stoichiometric ones. Altogether, structural characterizations by RHEED, STEM, and XRD provide almost identical results for $SrRu_{0.7}O_3$ and $SrRuO_3$, except for the subtle change in the lattice parameter.

We subsequently investigated the (magneto)transport properties of the Ru-deficient $SrRu_{0.7}O_3$ films, which were measured by a standard four-point-probe method with Ag electrodes deposited on the film surfaces without any additional processing. The distance between the two voltage electrodes was 2 mm. To begin with, the temperature dependence of the longitudinal resistivity $\rho_{xx}$ of the $SrRu_{0.7}O_3$ films with various thickness $t$ is shown in Fig. 3(a). No external magnetic field was applied. The $\rho_{xx}$ values of the films with $t = 5 - 60$ nm decrease with decreasing temperature $T$, indicating that these films are metallic in the whole temperature range measured. The $\rho_{xx}(T)$ curves converge in the thickness range of 10–60 nm—$\rho_{xx}$(300 K) ≈ 180 μΩ·cm and $\rho_{xx}$(2 K) ≈ 30 μΩ·cm, while the curve for the 5-nm-thick film shows an almost parallel shift to higher $\rho_{xx}$ by a few tens of μΩ·cm. These transport properties of the SrRu$_{0.7}$O$_3$ films with $t$ = 10–60 nm are in stark contrast to the stoichiometric SrRuO$_3$ case. In Fig. 3(b), $t$-dependent residual resistivity $\rho_{Res}$ [$\rho_{xx}$(2 K)] for the SrRu$_{0.7}$O$_3$ and SrRuO$_3$ films is shown. The data for the stoichiometric films are from our previous report.[19] The $\rho_{Res}$ of the Ru-deficient films stays almost constant for $t$ = 10–60 nm, whereas it gradually but monotonically increases from 2.0 to 7.5 μΩ·cm with decreasing thickness from 60 to 10 nm in the stoichiometric ones. Furthermore, $\rho_{Res}$ is higher in SrRu$_{0.7}$O$_3$ than in SrRuO$_3$ at each thickness. These results indicate the transport properties of the SrRu$_{0.7}$O$_3$ films are dominated by the scattering due to the Ru vacancies, rather than by interface-driven disorders, especially when $t$ ≥ 10 nm. In the stoichiometric films, an abrupt increase in $\rho_{Res}$ occurs in the thickness range of $t$ ≤ 10 nm, suggesting that carrier scattering at the interface-driven disorders plays a dominant role. Note that we consider interface-driven disorders whose types are insensitive to STEM observations. It is also noteworthy that SrRuO$_3$ might have been regarded as a bad metal based on experimental results for Ru-deficient specimens like SrRu$_{0.7}$O$_3$, since ten-times larger $\rho_{Res}$ leads to one order of magnitude smaller mean free path estimation.[8]

In Fig. 3(a), $\rho_{xx}(T)$ curves for the thinner films are also plotted. The $\rho_{xx}$ values of the 2-nm-thick film decrease with decreasing temperature from 300 K down to $T_{\rho min}$, at which the $\rho_{xx}$ becomes minimum, but for $T < T_{\rho min}$, $\rho_{xx}$ increases with decreasing temperature. This insulating behavior (d$\rho_{xx}$/d$T$ < 0) possibly stems from the weak-localization in the low-temperature range,[20,25–29] which is enhanced due to structural disorders, including lattice point defects (Ru vacancies) and some boundaries.[44] With a further decrease in $t$ down to 1 nm [inset in Fig. 3(a)], the $\rho_{xx}$ values become three to four orders of magnitude larger than those of the other films, and d$\rho_{xx}$/d$T$ is negative for the whole measurement temperature range (2–300 K); that is, the influence of interface-driven disorder prevails against Ru deficiencies in ultrathin films ($t$ ≤ 2 nm). Similar thickness-dependent insulating behaviors have been reported by many other groups.[20,25–29] In most of those studies, insulating behavior starts to appear when the film thickness is 2–4 nm,[25–28] implying that those films are subject to interface-driven disorders and possibly also to the cation off-stoichiometry problem. This interpretation seems reasonable as the high-crystalline-quality stoichiometric thin film with $t$ = 2 nm was metallic in our previous study.[19] In Fig. 3(c), we compare the $t$ dependence of the RRR [= $\rho_{xx}$(300 K)/$\rho_{xx}$(2 K)] between the SrRu$_{0.7}$O$_3$ and SrRuO$_3$ films with $t$ ≥ 2nm. The thickness dependence is substantially smaller in SrRu$_{0.7}$O$_3$ than in SrRuO$_3$, which is consistent with our consideration that the transport process is mainly limited by scattering within a lattice (Ru vacancies) for SrRu$_{0.7}$O$_3$ and interface-driven disorders for SrRuO$_3$.

When $t$ ≥ 5nm, the $\rho_{xx}(T)$ curves show clear kinks at around 140 K [arrow in Fig. 3(a)], at which the ferromagnetic transition occurs and spin-dependent scattering is suppressed.[8] To highlight the ferromagnetic transition, we plot the derivative resistivity $d\rho_{xx}/dT$ as a function of $T$ [Fig. 3(d)]. Here, we define $T_C$ as the temperature at which the $d\rho_{xx}/dT$ or $\rho_{xx}(T)$ curves show a clear peak or kink [solid arrows in Fig. 3(d)], respectively. We note that, in general, the $T_C$ determined from $d\rho_{xx}/dT$ is a few K lower than the values measured from the temperature dependence of the magnetization.[27] Apart from a small hump at higher temperatures observed for the films with $t$ = 40 and 60 nm, $d\rho_{xx}/dT$ curves and peak positions are essentially identical in the films with $t = 10-60$ nm, which exhibit a common $T_C$ of 140 K. The $t$ dependence of $T_C$ in Fig. 3(e), in which the $T_C(t)$ curves for SrRu$_{0.7}$O$_3$ and SrRuO$_3$ show a parallel shift, indicates that the highly Ru-deficient SrRu$_{0.7}$O$_3$ films inherently have a distinct $T_C$ of ≈ 140 K in the absent of interface-driven disorders ($t = 10-60$ nm), and that their $T_C$ values are 10 K lower than that of the stoichiometric SrRuO$_3$ films. In other words, the influence of the 30% Ru-deficiency on the magnetic ordering temperature is as small as 10 K. For $t < 10$ nm, $T_C$ sharply decreases (down to ≈ 100 K) with decreasing thickness irrespective of SrRu$_{0.7}$O$_3$ or SrRuO$_3$, indicating that the influence of the interface-driven disorders on the reduction of the exchange interaction (and thus the decrease in $T_C$) is larger than that of the Ru-deficiencies.

As mentioned above, the RRR of the SrRu$_{0.7}$O$_3$ films takes an almost constant value of about 5 when $t = 10-60$ nm and decreases to a value of a little larger than unity for $t = 2$ nm [Fig. 3(c)] (here, the residual resistivity for the 2-nm-thick film is estimated based on the $\rho_{min}$ value). On the other hand, $T_C$ stays almost constant (≈ 140 K) for $t = 10-60$ nm and sharply decreases with decreasing $t$ from 10 to 2 nm. As a result, $T_C$ sharply vary as a function of RRR for the Ru deficient SrRu$_{0.7}$O$_3$ films [Fig. 3(f)]. This trend for the SrRu$_{0.7}$O$_3$ films is similar to what is derived for stoichiometric SrRuO$_3$ films.[18,19] For the stoichiometric SrRuO$_3$ films, the RRR values varied in a much wider range; ultrahigh-quality SrRuO$_3$ films with $t = 63$ nm showed a very high RRR value of larger than 84.[19] The intermediate data points are obtained during the growth condition optimization processes and/or by varying the thickness of the SrRuO$_3$ films.[18,19] Accordingly, different RRR values have various origins, including Ru and/or O vacancies, orthorhombic domains, and interface-driven disorders.[8,30,31] Nevertheless, the universal trend in the $T_C$ vs. RRR plots indicate that $T_C$ is rather insensitive to the RRR when it is ≥ 20, irrespective of the origins of the lower RRR, whereas $T_C$ is very sensitive to the RRR when it is < 20 in the material system of SrRu$_{1-x}$O$_3$ including $x = 0$. The threshold value of around 20 coincides with what is required to observe transport phenomena specific to the magnetic Weyl semimetal state.[18,19]

We also performed magnetotransport measurements on the Ru-deficient SrRu$_{0.7}$O$_3$ films by the standard four-point-probe method. In Fig. 4(a), we show the $t$ dependence of the magnetoresistance (MR) [($\rho_{xx}(B)-\rho_{xx}(0$ T$))/\rho_{xx}(0$ T$)$], where the magnetic field $B$ is applied in the out-of-plane [001] direction of the SrTiO$_3$ substrate at 2 K. For comparison, in Ref. [19], the signatures of quantum transport of the Weyl fermions, e.g., the unsaturated linear positive MR accompanied by quantum oscillations having a π Berry phase, are observed in the stoichiometric SrRuO$_3$ films with $t ≥ 10$ nm. In the Ru-deficient SrRu$_{0.7}$O$_3$ films, however, the MR at high magnetic fields is negative and the SdH oscillations are not observed for any films, irrespective of the thickness [Fig. 4(a)]. From the results shown in Fig. 4(a), we extracted the thickness dependence of the MR ratios for the highest applied magnetic field [($\rho_{xx}(9$ T$)-\rho_{xx}(0$ T$))/\rho_{xx}(0$ T$)$] at 2 K [Fig. 4(b)]. The MR ratio of the SrRu$_{0.7}$O$_3$ films increases with increasing thickness and saturates to the MR = 0% line. In contrast to the stoichiometric SrRuO$_3$ case,[19] which is also plotted in Fig. 4(b), the thickness dependence is weak, and the values of the MR ratio never exceed zero. Since the positive MR comes from the Weyl fermions in SrRuO$_3$,[18,19] the magnetic Weyl semimetal state is not realized in the SrRu$_{0.7}$O$_3$ even when $t = 60$ nm; the Weyl semimetal state is observed only in the stoichiometric SrRuO$_3$ films with $t ≥ 10$ nm. These results indicate two scenarios: 1) the scattering at the Ru vacancies in the SrRu$_{0.7}$O$_3$ films hinders the emergence of quantum transport phenomena stemming from the Weyl fermions, or 2) the magnetic Weyl semimetal state itself is not realized in SrRu$_{0.7}$O$_3$ near the Fermi level *per se* due to the difference in electronic structure. The MR vs. RRR plot of the Ru-deficient SrRu$_{0.7}$O$_3$ films seems to fall within the trend derived from the stoichiometric SrRuO$_3$ films [Fig. 4(c)]. For SrRuO$_3$, the MR ratio increases with increasing RRR and eventually has positive values when the RRR is larger than 20, meaning that the Weyl fermions dominate the transport properties when the SrRuO$_3$ film has less defects.

In Fig. 4(a), for the SrRu$_{0.7}$O$_3$ films with $t \leq 10$ nm, anisotropic magnetoresistance (AMR),[45] which is proportional to the relative angle between the electric current and the magnetization, is clearly observed below 3 T at 2 K. In SrRuO$_3$, the AMR peak position corresponds to the coercive field $H_c$.[18] On the other hand, SrRu$_{0.7}$O$_3$ films with $t \geq 20$ nm show hysteresis in the MR above $H_c$ of the 10-nm-thick film (~1 T). Such features in the MR hysteresis suggest the existence of two different magnetic components in the SrRu$_{0.7}$O$_3$ films. The two component scenario is also supported by hump structures observed in the $d\rho_{xx}/dT$ vs. temperature curves for the films with $t = 40$ and 60 nm. The hump structures are discernible at around 150 K [dashed arrows in Fig. 3(d)], indicating the existence of two distinct $T_C$ values,[46] possibly due to inhomogeneous distribution of the Ru vacancies.

In summary, we investigated the structural and transport properties of the highly Ru-deficient SrRu$_{0.7}$O$_3$ thin films prepared by MBE on (001) SrTiO$_3$ substrates, and compared the results to those for stoichiometric SrRuO$_3$ thin films. To distinguish the influence of the two types of disorders—Ru vacancies within lattices and interface-driven disorders—SrRu$_{0.7}$O$_3$ thin films with various thicknesses ($t = 1 - 60$ nm) were prepared. It turned out that the influence of the former dominates the electrical and magnetic properties when $t \geq 5 - 10$ nm, while that of the latter does for $t \leq 5 - 10$ nm. Emphasis should be placed on the fact that the crystalline quality of the SrRu$_{0.7}$O$_3$ films is as high as that of ultrahigh-quality SrRuO$_3$ thin films, and that structural characterizations by RHEED, STEM, and XRD cannot distinguish the difference between SrRu$_{0.7}$O$_3$ and SrRuO$_3$ except for a subtle change in the lattice parameter. In view of the Curie temperature $T_C$, SrRu$_{0.7}$O$_3$ ($T_C \approx 140$ K) can be regarded as a material distinct from SrRuO$_3$ ($T_C \approx 150$ K). From the viewpoint of the transport properties, SrRu$_{0.7}$O$_3$ can be understood to be highly Ru-deficient SrRuO$_3$, where Ru vacancies serve as charge carrier scatter centers. Even if 30% of Ru sites are vacant, metallic conduction ($d\rho_{xx}/dT > 0$) is preserved for the films with $t \geq 5$ nm. This large tolerance in Ru composition is advantageous for some practical applications as an epitaxial conducting layer. At present, the ultrathin films ($t < 10$ nm) are inevitably subject to interface-driven disorders, irrespective of SrRu$_{0.7}$O$_3$ or SrRuO$_3$. Adopting different substrates that have less lattice mismatch may help to overcome the interface-driven disorder problem.

**AUTHORS' CONTRIBUTIONS**

Y.K.W. conceived the idea, designed the experiments, and led the project. Y.K.W. and Y.K. grew the samples. Y.K.W. and S.K.T. carried out the sample characterizations. S.K.T. and Y.K.W. carried out the transport measurements and analyzed the data. All authors contributed to the discussion of the data. Y.K.W. and S.K.T. co-wrote the paper with input from all authors.

**DATA AVAILABILITYY**

The data that support the findings of this study are available from the corresponding author upon reasonable request.

**References**

[1] J. J. Randall and R. Ward, J. Am. Chem. Soc. **81**, 2629 (1959).

[2] J. M. Longo, P. M. Raccah, and J. B. Goodenough, J. Appl. Phys. **39**, 1327 (1968).

[3] R. J. Bouchard and J. L. Gillson, Mater. Res. Bull. **7**, 873 (1972).

[4] L. Klein and J. S. Dodge, C. H. Ahn, J. W. Reiner, L. Mieville, T. H. Geballe, M. R. Beasley, and A. Kapitulnik, J. Phys. Condens. Matter **8**, 10111, (1996).

[5] C. B. Eom, R. J. Cava, R. M. Fleming, J. M. Phillips, R. B. van Dover, J. H. Marshall, J. W. P. Hsu, J. J. Krajewski, and W. F. Peck, Jr., Science **258**, 1766, (1992).

[6] X. D. Wu, S. R. Foltyn, R. C. Dye, Y. Coulter, and R. E. Muenchausen, Appl. Phys. Lett. **62**, 2434 (1993).

[7] M. Izumi, K. Nakazawa, Y. Bando, Y. Yoneda, and H. Terauchi, J. Phys. Soc. Jpn. **66**, 3893 (1997).

[8] G. Koster, L. Klein, W. Siemons, G. Rijnders, J. S. Dodge, C.-B. Eom, D. H. A. Blank, M. R. Beasley, Rev. Mod. Phys. **84**, 253 (2012).

[9] D. Kan, M. Mizumaki, T. Nishimura, and Y. Shimakawa, Phys. Rev. B, **94**, 214420, (2016).

[10] H. P. Nair, Y. Liu, J. P. Ruf, N. J. Schreiber, S.-L. Shang, D. J. Baek, B. H. Goodge, L. F. Kourkoutis, Z.-K. Liu, K. M. Shen, and D. G. Schlom, APL Mater. **6**, 046101 (2018).

[11] W. Siemons, G. Koster, A. Vailionis, H. Yamamoto, D. H. A. Blank, and M. R. Beasley, Phys. Rev. B **76**, 075126 (2007).

[12] J. Thompson, J. Nichols, S. Lee, S. Ryee, J. H. Gruenewald, J. G. Connell, M. Souri, J. M. Johnson, J. Hwang, M. J. Han, H. N. Lee, D.-W. Kim, and S. S. A. Seo, Appl. Phys. Lett. **109**, 161902 (2016).

[13] D. E. Shai, C. Adamo, D. W. Shen, C. M. Brooks, J. W. Harter, E. J. Monkman, B. Burganov, D. G. Schlom, and K.M. Shen, Phys. Rev. Lett. **110**, 087004 (2013).

[14] A. P. Mackenzie, J. W. Reiner, A. W. Tyler, L. M. Galvin, S. R. Julian, M. R. Beasley, T. H. Geballe, and A. Kapitulnik, Phys. Rev. B **58**, R13318 (1998).

[15] D. C. Worledge and T. H. Geballe, Phys. Rev. Lett. **85**, 5182 (2000).

[16] K. S. Takahashi, A. Sawa, Y. Ishii, H. Akoh, M. Kawasaki, and Y. Tokura, Phys. Rev. B **67**, 094413 (2003).

[17] Z. Li, S. Shen, Z. Tian, K. Hwangbo, M. Wang, Y. Wang, F. M. Bartram, L. He, Y. Lyu, Y. Dong, G. Wan, H. Li, N. Lu, J. Zang, H. Zhou, E. Arenholz, Q. He, L. Yang, W. Luo, and P. Yu, Nat. Commun. **11**, 184 (2020).

[18] K. Takiguchi, Y. K. Wakabayashi, H. Irie, Y. Krockenberger, T. Otsuka, H. Sawada, S. A. Nikolaev, H. Das, M. Tanaka, Y. Taniyasu, and H. Yamamoto, Nat. Commun. **11**, 4969 (2020).

[19] S. K. Takada, Y. K. Wakabayashi, Y. Krockenberger, S. Ohya, M. Tanaka, Y. Taniyasu, and H. Yamamoto, arXiv:2011.03670.

[20] H. Boschker, T. Harada, T. Asaba, R. Ashoori, A. V. Boris, H. Hilgenkamp, C. R. Hughes, M. E. Holtz, L. Li, D. A. Muller, H. Nair, P. Reith, X. R. Wang, D. G. Schlom,

A. Soukiassian, and J. Mannhart, Phys. Rev. X **9**, 011027 (2019).
21. Z. Cui, A. J. Grutter, H. Zhou, H. Cao, Y. Dong, D. A. Gilbert, J. Wang, Y.-S. Liu, J. Ma, Z. Hu, J. Guo, J. Xia, B. J. Kirby, P. Shafer, E. Arenholz, H. Chen, X. Zhai, and Y. Lu, Sci. Adv. **6**, eaay0114 (2020).
22. Y. Araki and K. Nomura, Phys. Rev. Appl. **10**, 014007 (2018).
23. C. H. Lee, S. Imhof, C. Berger, F. Bayer, J. Brehm, L. W. Molenkamp, T. Kiessling, and R. Thomale, Commun. Phys. **1**, 39 (2018).
24. S. M. Rafi-Ul-Islam, Z. B. Siu, and M. B. A. Jalil, Commun. Phys. **1**, 72 (2020).
25. D. Toyota, I. Ohkubo, H. Kumigashira, M. Oshima, T. Ohnishi, M. Lippmaa, M. Takizawa, A. Fujimori, K. Ono, M. Kawasaki, and H. Koinuma, Appl. Phys. Lett. **87**, 162508 (2005).
26. X. Shen, X. Qiu, D. Su, S. Zhou, A. Li, and D. Wu, J. Appl. Phys. **117**, 015307 (2015).
27. A. Rastogi, M. Brahlek, J. M. Ok, Z. Liao, C. Sohn, S. Feldman, and H. N. Lee, APL Mater. **7**, 091106 (2019).
28. J. Xia, W. Siemons, G. Koster, M. R. Beasley, and A. Kapitulnik, Phys. Rev. B **79**, 140407(R) (2009).
29. Y. J. Chang, C. H. Kim, S.-H. Phark, Y. S. Kim, J. Yu, and T. W. Noh, Phys. Rev. Lett. **103**, 057201 (2009).
30. Q. Gan, R. A. Rao, and C. B. Eom, Appl. Phys. Lett. **70**, 1962 (1997).
31. J.C. Jiang, W. Tian, X. Pan, Q. Gan, C.B. Eom, Mater. Sci. Eng. B **56**, 152 (1998).
32. Y. K. Wakabayashi, T. Otsuka, Y. Taniyasu, H. Yamamoto, and H. Sawada, Appl. Phys. Express **11**, 112401 (2018).
33. Y. K. Wakabayashi, T. Otsuka, Y. Krockenberger, H. Sawada, Y. Taniyasu, and H. Yamamoto, APL Mater. **7**, 101114 (2019).
34. M. Naito and H. Sato, Appl. Phys. Lett. **67**, 2557 (1995).
35. H. Yamamoto, Y. Krockenberger, and M. Naito, J. Cryst. Growth **378**, 184 (2013).
36. Y. K. Wakabayashi, Y. Krockenberger, N. Tsujimoto, T. Boykin, S. Tsuneyuki, Y. Taniyasu, and H. Yamamoto, Nat. Commun. **10**, 535 (2019).
37. D. E. Newbury and N. W. M. Ritchie, J. Mater. Sci. **50**, 493 (2015).
38. P. A. Ferreiros, P. R. Alonso, D. P. Quiros, E. Zelaya, G.H. Rubiolo, Journal of Alloys and Compounds 847, 156372 (2020).
39. R. Kotz, H. J. Lewerenz, and S. Stucki, J. Electrochem. Soc. **130**, 825 (1983).
40. M. Takizawa, D. Toyota, H. Wadati, A. Chikamatsu, H. Kumigashira, A. Fujimori, M. Oshima, Z. Fang, M. Lippmaa, M. Kawasaki, and H. Koinuma, Phys. Rev. B **72**, 060404(R) (2005).
41. F. Yan, M.-O. Lai, L. Lu, and T.-J. Zhu, J. Phys. Chem. C **114**, 6994 (2010).
42. G. Berti, S. Sanna, C. Castellano, J. V. Duijn, R. R. Bustos, L. Bordonali, G. Bussetti, A. Calloni, F. Demartin, L. Duò, and A. Brambilla, J. Phys. Chem. C **120**, 11763 (2016).
43. B. Dabrowski, O. Chmaissem, P. W. Klamut, S. Kolesnik, M. Maxwell, J. Mais, Y. Ito, B. D. Armstrong, J. D. Jorgensen, and S. Short, Phys. Rev. B **70**, 014423 (2004).
44. G. Herranz, F. Sanchez, B. Martınez, J. Fontcuberta, M. V. Garcıa-Cuenca, C. Ferrater, M. Varela, and P. Levy, Eur. Phys. J. B **40**, 439 (2004).
45. D. B. Kacedon, R. A. Rao, and C. B. Eom, Appl. Phys. Lett. **71**, 1724 (1997).
46. J. Lu, L. Si, X. Yao, C. Tian, J. Wang, Q. Zhang, Z. Lai, I. A. Malik, X. Liu, P. Jiang, K. Zhu, Y. Shi, Z. Luo, L. Gu, K. Held, W. Mi, Z. Zhong, C-W. Nan, and J. Zhang, Phys. Rev. B **101**, 214401 (2020).

**Figures and figure captions**

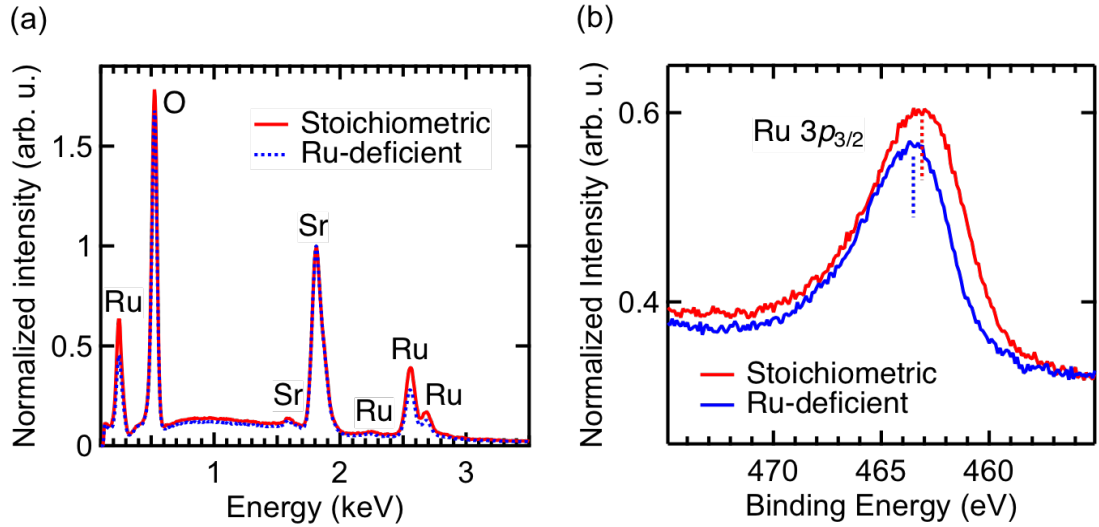

FIG. 1. (a) EDS spectra of the stoichiometric and Ru-deficient SrRuO$_3$ films with $t = 60$ nm, which were taken from a wide area (85 × 115 $\mu$m$^2$) of the films. (b) Ru $3p_{3/2}$ XPS spectra of the stoichiometric and Ru-deficient SrRuO$_3$ films with $t = 60$ nm. Dashed lines indicate peak positions. The XPS spectra are normalized at the Sr $3d_{5/2}$ XPS peak intensities for easy comparison.

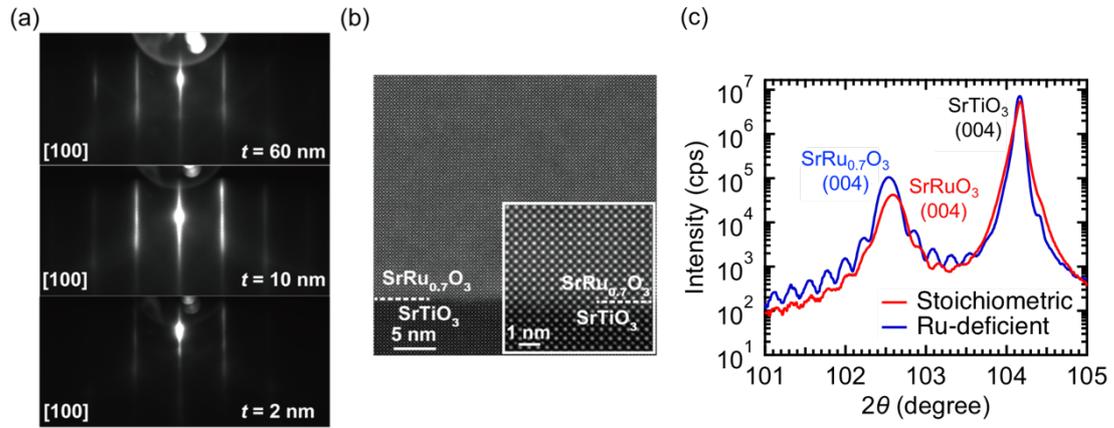

FIG. 2. (a) RHEED patterns of the Ru-deficient SrRu$_{0.7}$O$_3$ films with $t$ = 60, 10 and 2 nm taken along the [100] axis of the SrTiO$_3$ substrates. (b) HAADF-STEM images of the Ru-deficient SrRu$_{0.7}$O$_3$ films taken along the [100] axis of the SrTiO$_3$ substrates. The inset in (b) is the magnified image near the interface. (c) Comparison of X-ray diffraction patterns between the SrRu$_{0.7}$O$_3$ and SrRuO$_3$ thin films with $t$ = 60 nm.

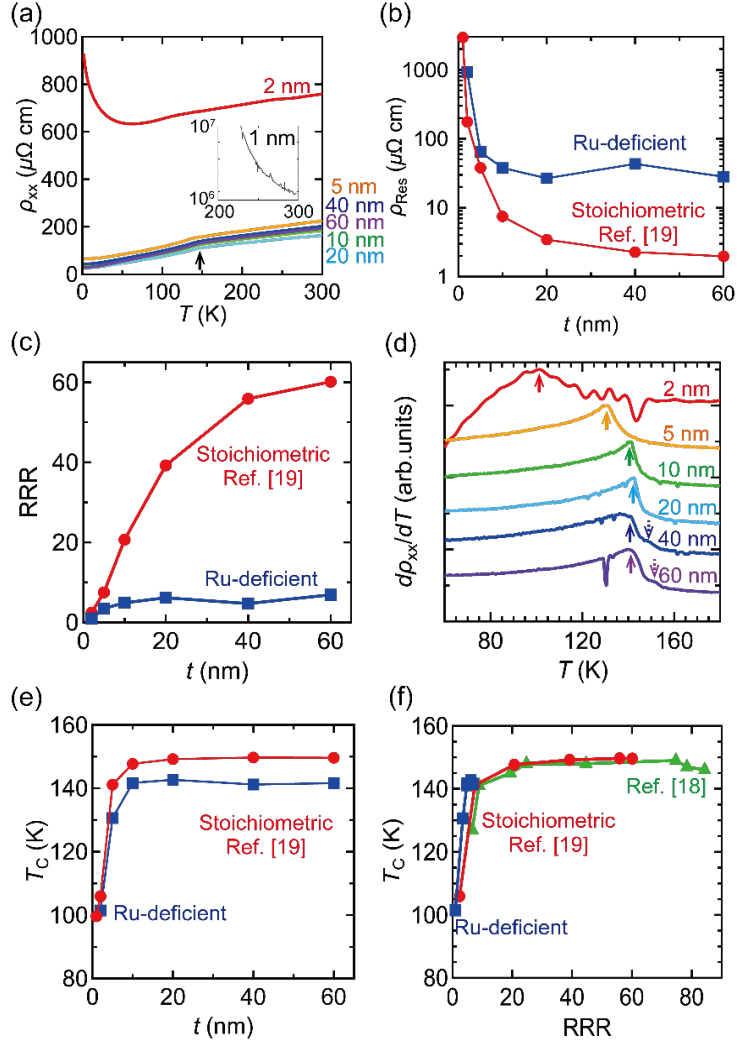

FIG. 3. (a) Temperature dependence of the resistivity $\rho_{xx}$ of the Ru-deficient $SrRu_{0.7}O_3$ films with various thicknesses $t$ (=1–60 nm). The arrow indicates the kinks. The inset in (a) shows the data for $t$ = 1 nm. (b) Thickness $t$ dependence of the residual resistivity $\rho_{Res}$ of the Ru-deficient $SrRu_{0.7}O_3$ films. (c) Thickness $t$ dependence of the RRR. The data for the stoichiometric $SrRuO_3$ films reported in Ref. [19] are also plotted. (d) Temperature dependence of the differential resistivity $d\rho_{xx}/dT$ of the Ru-deficient $SrRu_{0.7}O_3$ films for various $t$ (=2–60 nm). Arrows indicate the peak or kink positions in the $d\rho_{xx}/dT$ curves. Dashed arrows indicate shoulder structures. In (d), the $d\rho_{xx}/dT$ has been offset for easy viewing. (e) Thickness $t$ dependence of the $T_C$ obtained from the peak positions of the differential resistivity $d\rho_{xx}/dT$ in (d). The data for the stoichiometric $SrRuO_3$ films reported in Ref. [19] are also plotted. (f) RRR dependence of the $T_C$ of the Ru-deficient $SrRu_{0.7}O_3$ films. The data reported in Refs. [18] and [19] are also plotted.

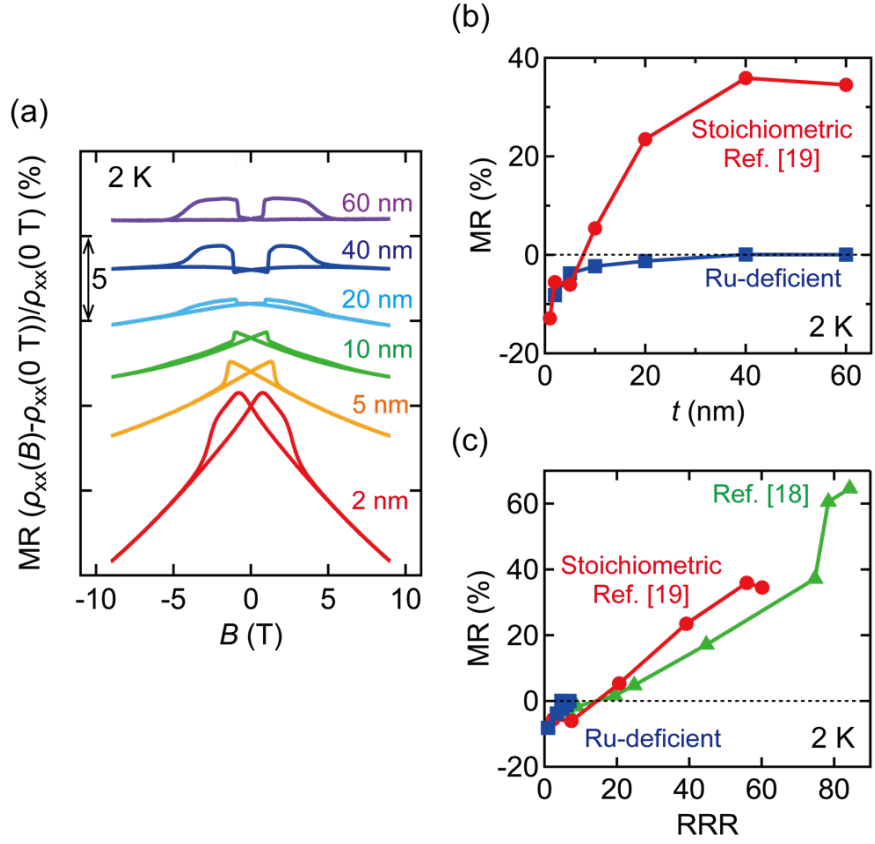

FIG. 4. (a) Thickness $t$ dependence of MR $(\rho_{xx}(B)-\rho_{xx}(0\,\mathrm{T}))/\rho_{xx}(0\,\mathrm{T})$ of the Ru-deficient SrRu$_{0.7}$O$_3$ films at 2 K with $B$ applied in the out-of-plane [001] direction of the SrTiO$_3$ substrate. (b), (c) Thickness $t$ (b) and RRR (c) dependence of the MR ratio at 2 K with $B$ = 9 T applied in the out-of-plane [001] direction of the SrTiO$_3$ substrate.